\documentclass[11pt,twocolumn]{article}
\pdfoutput=1
\usepackage{authblk,natbib,amsmath,pgf,vmargin,float,nnfootnote}
\usepackage[font=scriptsize]{caption}
\setpapersize{A4}
\setmarginsrb{25mm}{25mm}{25mm}{25mm}{0cm}{0cm}{0cm}{0cm} 

\usepackage[compact]{titlesec}
\titlespacing{\subsection}{0pt}{.15cm}{.15cm}
\titlespacing{\subsubsection}{0pt}{.15cm}{.15cm}

\title{\vspace{-.9cm}
  SUPER-DROPLET APPROACH TO SIMULATE PRECIPITATING TRADE-WIND CUMULI -- 
  COMPARISON OF MODEL RESULTS WITH RICO AIRCRAFT OBSERVATIONS
}
\date{\vspace{-1.3cm}}

\author[1]{Sylwester Arabas\footnote{Correspondence to: Sylwester Arabas, Pasteura 7,\\ 02-093 Warsaw, Poland. E-mail: sarabas@igf.fuw.edu.pl}}
\author[2]{Shin-ichiro Shima\footnote{Affiliation at the time of research: Japan Agency for Marine-Earth Science and Technology, Kanagawa, Japan\\
\hrule
\noindent Paper presented at the 16\textsuperscript{th} International Conference on Clouds and Precipitation ICCP-2012, Leipzig, Germany\\
}}

\affil[1]{Institute of Geophysics, Faculty of Physics, University of Warsaw, Poland}
\affil[2]{Graduate School of Simulation Studies, University of Hyogo, Kobe, Japan} 

\begin{document}
  \maketitle\thispagestyle{empty}

  \subsection{INTRODUCTION}

  In this study we present a series of Large Eddy Simulations (LES) employing the Super-Droplet Method (SDM)
    for representing aerosol, cloud and warm-rain microphysics \citep{Shima_2008,Shima_et_al_2009}.
  SDM is a particle-based and probabilistic Monte-Carlo type model.
  The particle-based formulation helps to overcome the problem of parameterisation of processes occurring 
    at single-particle scale (micro- to millimetres) as source/sink terms in the LES equations solved on
    grids with cell dimensions of the order of tens of metres.
  The aim of this paper is to showcase the capabilities and point out the limitations of SDM in context of
    simulation of a field of precipitating clouds.

  \subsection{THE SUPER-DROPLET METHOD}

  The framework of the method consists of two mutually coupled simulation components:
    (i)~a~fluid flow solver computing (in an Eulerian sense) the evolution of fluid velocity field and evolution of
    the thermodynamic scalar quantities, and (ii)~a~particle-tracking logic computing (in a Lagrangian sense) 
    evolution of physical coordinates and physicochemical properties of a population of particles.
  The coupling between the Eulerian and Lagrangian components is bi-directional.
  The Lagrangian component feeds on the fluid velocity field in order to update the positions of super-droplets,
    and the thermodynamic fields to compute the condensational growth or evaporation rates.
  The Eulerian component feeds on the water vapour and heat source/sink rates resulting from condensation and 
    evaporation of water on the particles.
  Although mutually coupled, the Eulerian and Lagrangian computations are decomposed, in the sense that they
    happen sequentially and the state vectors of one component are constant from the standpoint
    of the other (within a time-step).

  The model does not differentiate between aerosol particles, cloud droplets, drizzle or rain drops.
  Each particle in the model (referred to as super-droplet) represents a multiplicity of real-world 
    particles of the same size and of the same chemical composition.
  From the standpoint of the Lagrangian component, the super-droplets are subject to three processes besides
    advection: (i) gravitational settling, (ii) condensational growth/evaporation and (iii) collisional growth.
  Consequently, the model covers representation of such cloud-microphysical processes as: cloud condensation nuclei
    (CCN) activation; drizzle formation by collisions of cloud droplets (autoconversion); accretion of cloud droplets
    by drizzle drops and raindrops, as well as coalescence of these larger hydrometeors (self-collection);
    and precipitation of drizzle and rain including aerosol wet deposition.
  Analogous particle-based techniques were recently used in context of simulations of precipitation-forming clouds e.g. by
    \citet{Andrejczuk_et_al_2010} and \citet{Franke_et_al_2011}; what distinguishes SDM however, 
    is the probabilistic Monte-Carlo type representation of the particle coalescence process.
  Probabilistic representation of particle coalescence has also been recently used in atmospheric modelling
    studies; however, to authors' knowledge, none of these models were bi-directionally coupled with LES --
    see e.g. \citet{Jensen_et_al_2008} for description of an adiabatic parcel model with Monte-Carlo coalescence
    or \citet{Riemer_et_al_2009} for description of particle-resolved aerosol transport model with Monte-Carlo
    coalescence but working off-line from the flow dynamics).

  \begin{table*}[t]\footnotesize
    \caption{\footnotesize
      List of model runs discussed in the paper.
      The run label denotes whether bulk (blk) or SDM (sdm) microphysics was used,
        as well as which grid resolution (coarse, middle or high) and super-droplet 
        number density was chosen.
      Coarse resolution corresponds to a quarter of the domain from the original RICO set-up 
        (i.e. grid box size of 100$\times$100$\times$40~m with 64$\times$64$\times$100 grid points);
        the middle and high resolutions denote settings resulting in halved and quartered 
        grid box dimensions, respectively (with the domain size kept constant).
      For each simulation there are five time-steps defined: long and short time-step of the Eulerian
        component (the short one used for sound-wave terms), the time-step used for integrating the
        condensational growth/evaporation equation, the time-step
        used for solving collisional growth using the Monte-Carlo scheme, and the time-step for integration 
        of particle motion equations.
    }\label{table}
    \vspace{-.2cm}
    \begin{center}
      \begin{tabular}{l|r|r|r|r|c}
        run label      & grid                        & dx=dy     & dz & time-steps [s]       & sd density [cm$^{-3}$] \\ \hline
        blk-coarse     & 64 $\times\,$64~$\times$100   & 100m       & 40m & 1.00/0.100 ~~~~~~n/a~~~~~~   & n/a                    \\ 
        sdm-coarse-8   & 64 $\times\,$64~$\times$100   & 100m       & 40m & 1.00/0.100/0.25/1.0/1.0   & 2.0$\times$10$^{-11}$  \\ 
        sdm-coarse-32  & 64 $\times\,$64~$\times$100   & 100m       & 40m & 1.00/0.100/0.25/1.0/1.0   & 8.0$\times$10$^{-11}$  \\ 
        sdm-coarse-128 & 64 $\times\,$64~$\times$100   & 100m       & 40m & 1.00/0.100/0.25/1.0/1.0   & 3.2$\times$10$^{-10}$  \\ 
        sdm-coarse-512 & 64 $\times\,$64~$\times$100   & 100m       & 40m & 1.00/0.100/0.25/1.0/1.0   & 1.3$\times$10$^{-09}$  \\ 
        sdm-middle-8   & 128$\times$128$\times$200   &  50m       & 20m & 0.50/0.050/0.25/1.0/1.0  & 1.6$\times$10$^{-10}$  \\ 
        sdm-middle-32  & 128$\times$128$\times$200   &  50m       & 20m & 0.50/0.050/0.25/1.0/1.0  & 6.4$\times$10$^{-10}$  \\ 
        sdm-middle-128 & 128$\times$128$\times$200   &  50m       & 20m & 0.50/0.050/0.25/1.0/1.0  & 2.6$\times$10$^{-09}$  \\ 
        sdm-high-8     & 256$\times$256$\times$400   &  25m       & 10m & 0.25/0.025/0.25/1.0/0.5   & 1.3$\times$10$^{-09}$  \\ 
      \end{tabular}
    \end{center}
    \vspace{-.8cm}
  \end{table*}

  \subsection{MODELLING SET-UP}

  All simulations in the present study were carried out using the Nagoya University 
    Cloud-Resolving Storm Simulator \citep[CReSS:][chapter~9.2 and references therein]{Tsuboki_2008}.
  CReSS is a non-hydrostatic compressible flow LES-type solver.
  Two types of moist processes representations were used: the SDM and a bulk microphysics
    model \citep[Kessler-type parameterisation implemented following][section 2b]{Klemp_et_al_1978}.
  The simulations are carried out using a set-up based on the RICO composite case defined in \citet{vanZanten_et_al_2011}
    and corresponding to atmospheric state measured and modelled in context of the Rain in Cumulus over Ocean
    (RICO) field project \citep{Rauber_et_al_2007}.
  The set-up defines initial profiles of potential temperature, specific humidity and wind
    characteristics for the trade-wind boundary layer following RICO observations \citep[see][]{Nuijens_et_al_2009}.
  This set-up was previously used in several modelling studies including: \citep{Stevens_et_al_2008, 
    Jiang_et_al_2009, Seifert_et_al_2010, Matheou_et_al_2011, Grabowski_et_al_2011}.
  Here the only exception from the original set-up is the domain size -- a quarter of the original domain was used
    (i.e. 6.4$\times$6.4$\times$4~km instead of 12.8$\times$12.8$\times$4~km), and the grid cell sizes -- several
    settings were used.
  Tests with full domain size (not discussed herein) revealed that quartering the domain enlarges the fluctuations in time 
    of cloud macroscopic properties; however, not to a level significant for the presented discussion.

  For simulations using SDM, the initial coordinates and sizes of particles are chosen randomly 
    with uniform distribution in physical space and bi-modal lognormal distribution in particle-size
    space \citep[size spectrum as specified in][section 2.2.3]{vanZanten_et_al_2011}.
  All particles are initially put in equilibrium with ambient humidity assuming 
    that all are composed of ammonium sulphate solution.
  The mean number of super-droplets per LES grid box varied from 8 to 512 in different simulations (see Table~\ref{table}).
  In all calculations discussed in the paper the probability of collisions and coalescence was defined following
    \citet[][section 3d, and references therein]{Hall_1980} hence no effects of small-scale turbulence 
    on the drop collision efficiency was taken into account.

  \begin{figure*}[hp]
    \noindent\center\includegraphics[width=\textwidth]{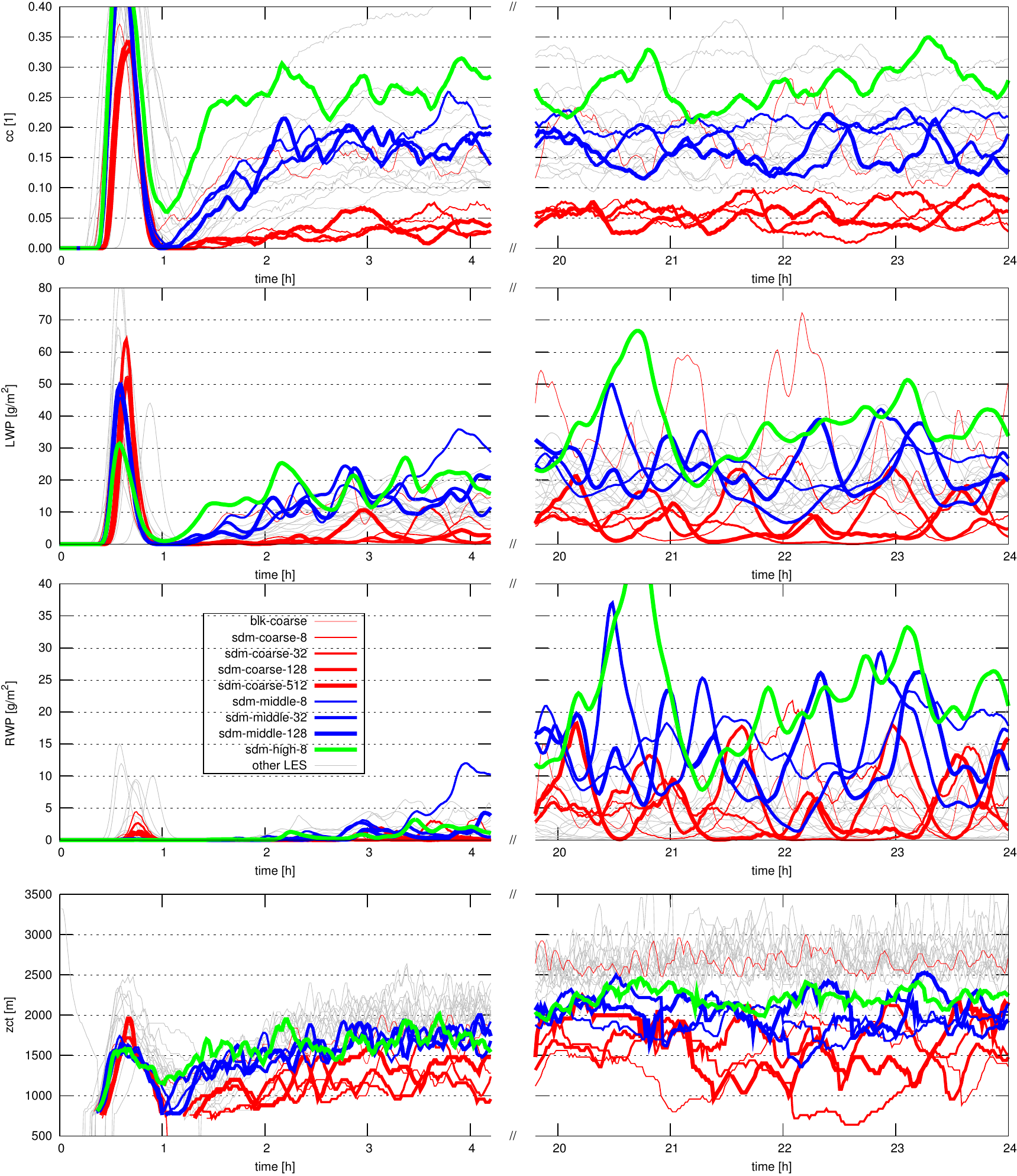}\\
    \caption{
      Time-series of cloud macrostructure characteristics defined following
        \citet[][section 2.4.2 and Table~4 therein]{vanZanten_et_al_2011}:
      (i) cloud cover ''cc'' is defined as the fraction of columns with at least one cloudy grid cell
        (a grid cell is defined as cloudy if the cloud water mixing ratio exceeds $0.01$~g~kg\textsuperscript{-1});
      (ii) liquid water path ''LWP'' is the mean over all columns of column-integrated liquid water content (density of both cloud and rain water);
      (iii) rain water path ''RWP'' is calculated in the same manner using rain water only;
      (iv) the cloud top height ''zct'' is the height of the top of the highest cloudy grid cell.
      For SDM simulations the cloud and rain water mixing ratios are diagnosed by summing over the population of particles 
        with radius smaller and larger than 40~$\mu m$, respectively.
      In the foreground there are plots depicting data from the nine simulations listed in Table~\ref{table}.
      In the background (plotted with thin grey lines) there are the results from LES simulations
        described in \citet[][data obtained at: http://knmi.nl/samenw/rico/]{vanZanten_et_al_2011}.
      See section~\ref{sec:macro} for discussion.
    }\label{scalars}
  \end{figure*}

  \subsection{SIMULATION RESULTS}

  A list of model runs and their corresponding labels used in the paper is given in Table~\ref{table}.
  The SDM simulations were carried with different grid resolutions and different super-droplet number
    densities.
  The table also lists different time-steps used in the simulations - in SDM the time-steps used for the 
    Eulerian and Lagrangian computations differ as the corresponding numerical stability constraints differ.

  Following the methodology of \citet{vanZanten_et_al_2011} the analyses 
    are restricted to the last four hours of the 24-hour model runs.
  Presented results are based on the LES grid data output every minute simulated time and
    super-droplet data (particle positions and sizes) output every 10 minutes simulated time.

  \subsubsection{CLOUD MACROSTRUCTURE}\label{sec:macro}

  Figure~\ref{scalars} presents time-series of scalar quantities characterising domain-wide 
    cloud field features: cloud cover (cc), liquid water path (LWP), rain water path (RWP)
    and cloud top height (zct).
  All quantities are labelled and defined following \citet[][section 2.4.2 and Table~4 therein, and caption of
    Figure~\ref{scalars} herein]{vanZanten_et_al_2011}.
  Presented time-series cover the first and the last four hours of simulations corresponding
    to the spin-up, and the relatively steady-state stages of the simulation, respectively.
  The figure is intended for comparison with analogous plots from modelling studies employing the RICO set-up:
    Fig.~2~in \citet{Stevens_et_al_2008},  Fig.~5~in \citet{Seifert_et_al_2010}, 
    Fig.~3~in \citet{vanZanten_et_al_2011}, Figs 1-3,7~in \citet{Matheou_et_al_2011} and
    Fig.~6~in \citet{Grabowski_et_al_2011}.

  The cloud-cover (cc) plots reveal considerable dependence of cc on the choice of grid
    resolution, with a significant increase of cloud cover for higher-resolution simulations.
  This is, at least partially, caused by the definition of cc which includes an arbitrary threshold value
    for the cloud water mixing ratio (e.g. in a hypothetic 2x2 chessboard water content distribution 
    with cc of $50$\%, downsampling to a single grid point could result in a zero cc).
  Furthermore, since the formation of convective clouds is triggered by vertical air motion and
    since refinement of the grid resolution helps to resolve in more detail the dynamics, 
    an increase in the number of (smaller) clouds may be expected \citep[][Sect.~3.2]{Matheou_et_al_2011}.

  \begin{figure*}[ht]
    \noindent\center\includegraphics[width=\textwidth]{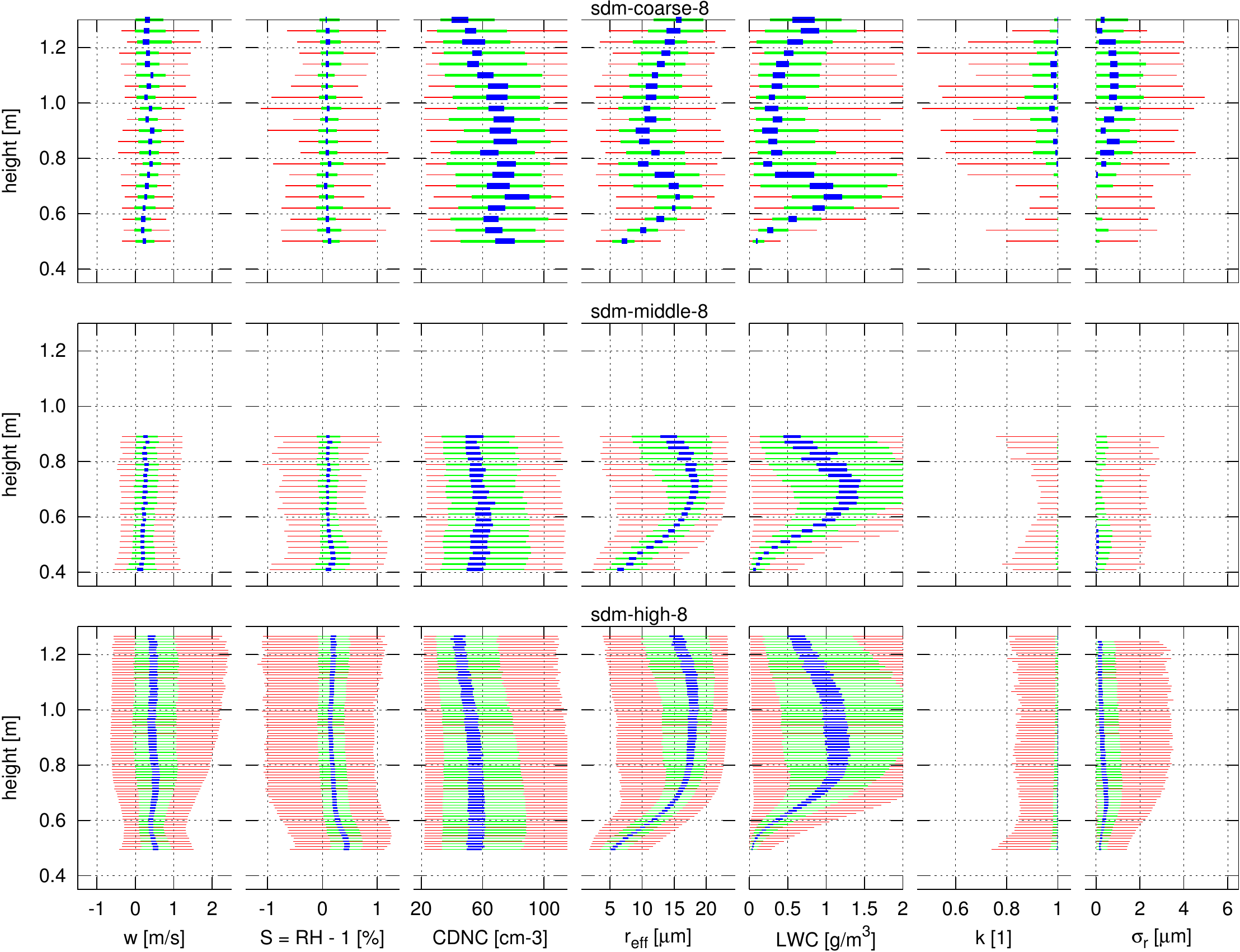}\\
    \caption{
      Height-resolved statistics of cloud droplet spectrum parameters derived from SDM simulation data.
      All plots in a row share the same y-axis representing height above sea level.
      See section~\ref{sec:fssp} for discussion and plot construction method description.
    \vspace{-.2cm}
    }\label{fssp1}
  \end{figure*}

  \begin{figure*}[ht]
    \noindent\center\includegraphics[width=\textwidth]{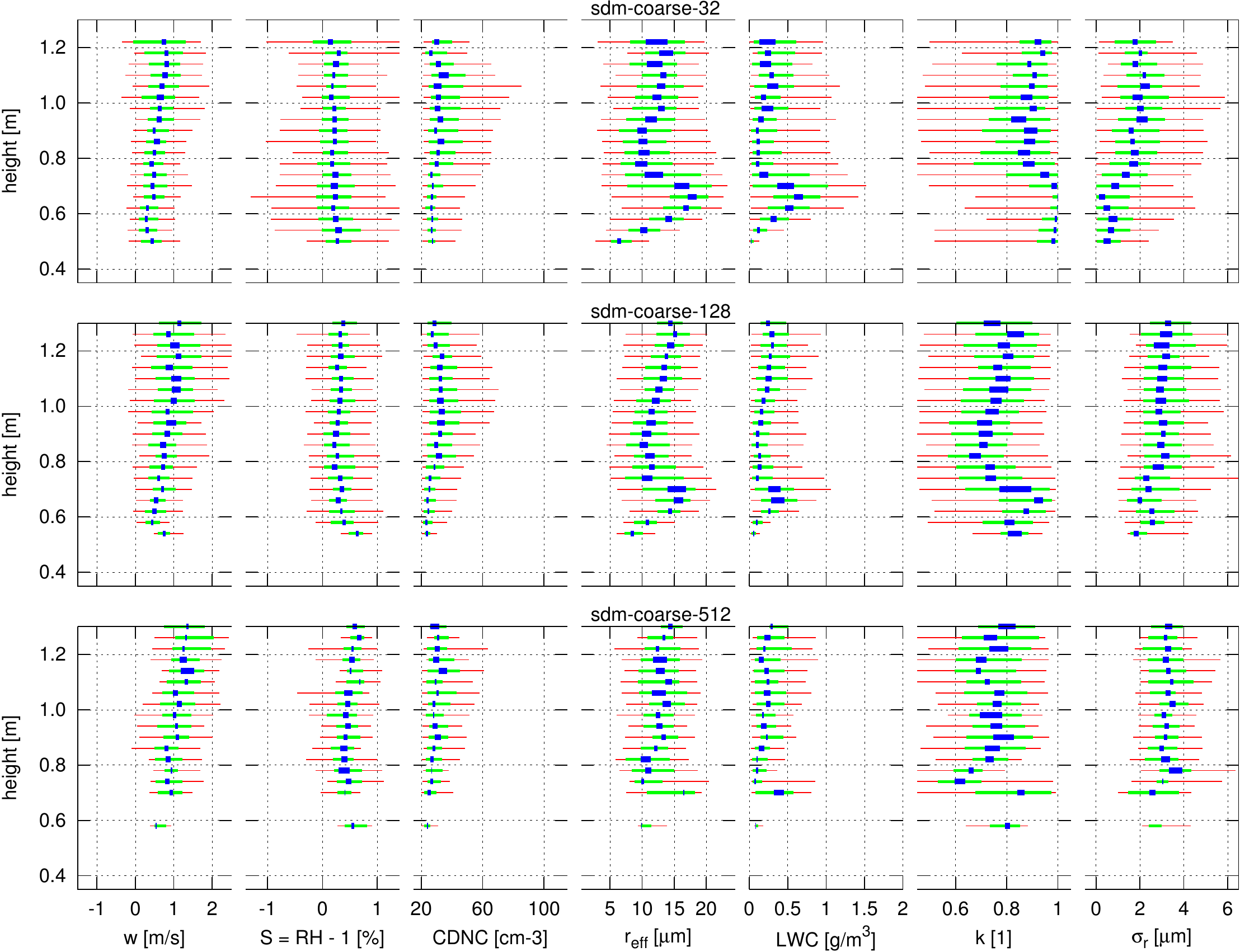}\\
    \caption{
      Same as Fig.~\ref{fssp1} for model runs: sdm-coarse-32, sdm-coarse-128 and sdm-coarse-512.
    \vspace{-.2cm}
    }\label{fssp2}
  \end{figure*}

  \begin{figure*}[ht]
    \noindent\center\includegraphics[width=\textwidth]{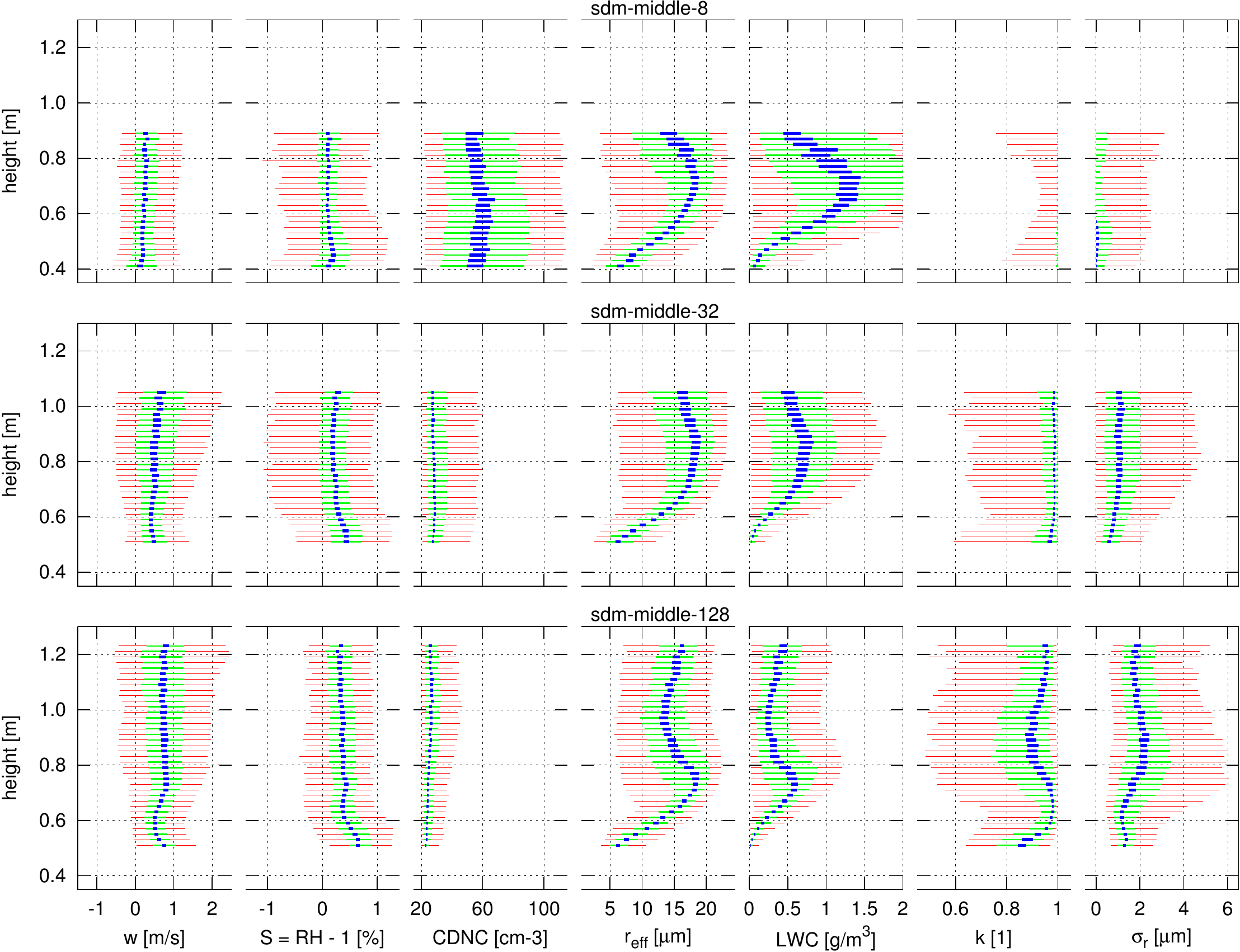}\\
    \caption{
      Same as Fig.~\ref{fssp1} for model runs: sdm-middle-8, sdm-middle-32 and sdm-middle-128.
    \vspace{-.2cm}
    }\label{fssp3}
  \end{figure*}

  Liquid water path (LWP) plots show that the increase of cc with increased resolution is
    correlated with increase of LWP.
  The values of LWP obtained in middle-resolution SDM simulations (blue lines) fall within the range 
    of values obtained with other LES from the study of \citet[][grey lines in Fig.~\ref{scalars} herein]{vanZanten_et_al_2011} for most of the simulated time.
  The high-resolution run (green line) features highest fluctuations of values of both
    LWP and RWP which likely corresponds to the low super-droplet density, and
    hence the low spectral resolution of the microphysics model.

  The rain water path was one of the quantities that varied most from one simulation to another
    in the study of \citet[][Sect.~3.2]{vanZanten_et_al_2011}.
  Similarly, the hereby presented results exhibit both significant variations in RWP with simulation time as
    well as significant departures from one simulation to another (i.e. when changing grid resolution
    or super-droplet density).
  A~notable feature is the high correlation of RWP with both LWP and cc.
  
  The plots of cloud-top evolution with time clearly show that the SDM-simulated cloud field
    is shallower than the one obtained with bulk microphysics or in other LES.
  While the employed definition of cloud-top height does not allow direct comparisons
    with measurements, arguably the SDM-predicted heights for middle- and high-resolution
    seem more comparable to those observed during RICO \citep[e.g.][Fig.~1]{Rauber_et_al_2007, Genkova_et_al_2007}.
  In general, the significant inhibition of convection in all coarse resolution runs suggest that
     the $100\times100\time40$ meter resolution is not enough when coupled with a Lagrangian
     model which, among other factors, relies on LES-predicted supersaturation field
     directly coupled with the vertical motions.

  \subsubsection{CLOUD MICROSTRUCTURE}\label{sec:fssp}

  \begin{figure*}[ht]
    \noindent\center\includegraphics[width=.8\textwidth]{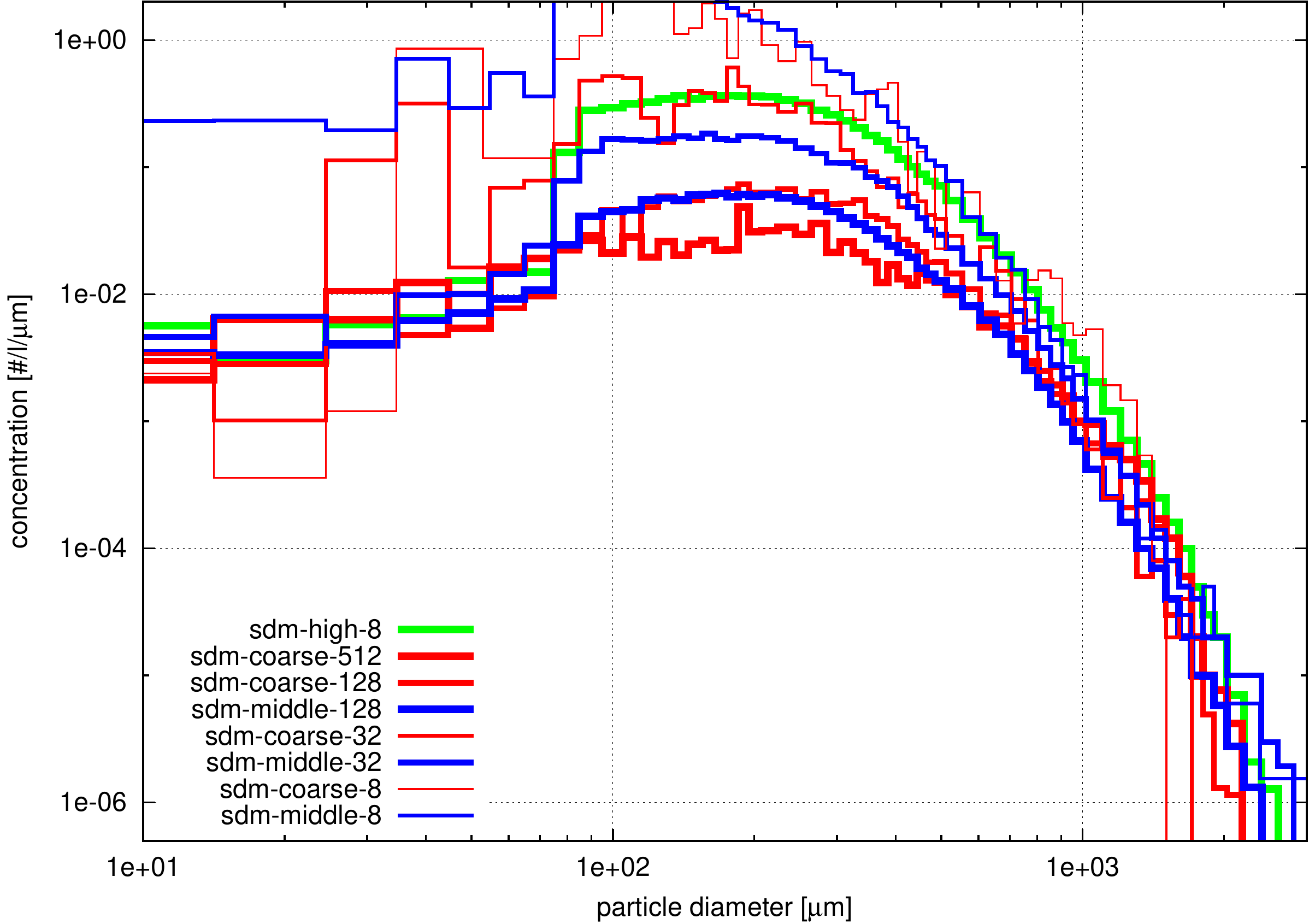}\\
    \caption{
      Comparison of model-predicted spectra for particles within size range of the OAP-2DS instrument,
        within grid cells with rain water mixing ratio $q_r>0.001$~g/kg and within
        an altitude range of $183\pm100$~m.
      Figure intended for comparison with Fig.~4 in \citet{Baker_et_al_2009}, see~section~\ref{sec:oap} herein for details.
      \vspace{-.2cm}
    }\label{oap}
  \end{figure*}

  Figures \ref{fssp1},\ref{fssp2} and~\ref{fssp3} present height-resolved statistics of
    the vertical velocity w, the supersaturation S, cloud droplet concentration CDNC, 
    droplet effective radius $r_{e\!f\!f}\!\!=<\!\!r^3\!\!>\!/\!<\!\!r^2\!\!>$, 
    liquid water content LWC, the cubed ratio of mean volume radius to effective 
    radius $k\!\!=<\!\!r^3\!\!>\!/\,r_{e\!f\!f}^3$, and the standard deviation of cloud droplet radius $\sigma_r$.
  The plots are intended for comparison with the analysis presented in \citet[][Figs.~1~and~2]{Arabas_et_al_2009}
    where the data from aircraft measurements during the RICO campaign using the Fast-FSSP optical cloud droplet 
    spectrometer \citep{Brenguier_et_al_1998} were analysed.
  The herein analysis of SDM simulation data is constrained to in-cloud regions defined as the grid boxes having 
    CDNC$>20$~cm\textsuperscript{-3} where CDNC is derived by summing over the super-droplets representing 
    particles of radius between 1 and 24 micrometers.
  The choices of the CDNC threshold and the spectral range correspond to those 
    characteristic of the Fast-FSSP probe.

  Plot construction method was chosen following the methodology of \citet{Arabas_et_al_2009}.
  For each level of the model grid and each plotted parameter 
    a list of values matching the in-cloud criterion is constructed, sorted and linearly interpolated
    to find the 5\textsuperscript{th}, 25\textsuperscript{th}, 45\textsuperscript{th}, 55\textsuperscript{th}, 
    75\textsuperscript{th} and 95\textsuperscript{th} percentiles.
  The lists are constructed from the LES-grid values (w, S) or super-droplet statistics calculated for each grid cell
    (CDNC, $r_{e\!f\!f}$, LWC, k and $\sigma_r$).
  The sample volumes are therefore defined by grid cell sizes and are of the order of $10^4$ -- $10^5\,m^3$, 
    while the sample volume for the $10\,Hz$ Fast-FSSP dataset used in \citet{Arabas_et_al_2009} is of the order 
    of $10^{-6}\,m^3$ -- a thing one has to bare in mind.
  The vertical extent of the measurement sample volume of ca. $10\,m$ is comparable to the grid cell size, though.
  Levels with less than $0.1 \cdot N_{max}$ data-points, where $N_{max}$ is the number of data-points 
    at the level with the highest number of data-points, are discarded from the analysis.
  Finally, the 5\textsuperscript{th} -- 95\textsuperscript{th} percentile, the interquartile, and the 45\textsuperscript{th} -- 55\textsuperscript{th} percentile ranges are
    plotted as a function of height using red, green and blue bars, respectively.
  The profiles composed of the 45\textsuperscript{th} 55\textsuperscript{th} percentile range bars are referred to as median profiles hereinafter.

  \paragraph{VERTICAL VELOCITY}

  Profiles of the vertical velocity generally show a gradual increase of the spread of
    values with the increasing grid resolution.
  It is consistent with the fact that the grid acts as a spatial filter in the LES.
  The median profiles resemble each other regardless of the resolution suggesting that
    the macroscale dynamics of the simulated cloud layer are robust to both the LES grid choice and 
    the super-droplet density choice (however, the differences in cloud-top heights plotted in Fig.~\ref{scalars}
    and evident in Figs.~\ref{fssp1}--\ref{fssp3} reveal that this robustness is limited).
  The slight increase of the vertical velocities with height visible in some of the profiles,
    and noted in the RICO observations analysis \citep[][Table.~2]{Gerber_et_al_2008}, may 
    correspond to the fact that only the more vigorous updraft regions were able to produce
    clouds reaching the upper part of the cloud field.

  \paragraph{SUPERSATURATION}

  The profiles of supersaturation, especially for the high and middle resolution,
    show the characteristic cloud-base maxima.
  This confirms that the model formulation, and the time resolution used allows to capture
    the influence of CCN activation kinetics on the evolution of supersaturation.
  Both the values of maximal supersaturation (ca. 1\%), as well as a roughly estimated
    supersaturation relaxation time (ca. $100$s assuming $\approx1$~m/s
    vertical velocity, and an $\approx100$m height scale over which the supersaturation
    falls off to an asymptotic value) correspond to those reported 
    in modelling studies that employed supersaturation-predicting models \citep[e.g.][]{Morrison_et_al_2008, Khvorostyanov_et_al_2008}.
  Comparison of the supersaturation prediction in the model with measurements is not viable
    as direct measurements of supersaturation in clouds are virtually unavailable \citep{Korolev_et_al_2003}.
  In none of the runs there is any other maximum of supersaturation visible along the profile.
  This suggests that the employed Lagrangian technique for representing water condensate inhibits the spurious
    production of cloud-edge supersaturation inherent in Eulerian models.
    \citep[see Sec.~1 in][and references therein]{Grabowski_et_al_2008}.
  The reason for it is likely \citep[as noted as well in][par.~10]{Andrejczuk_et_al_2008}
    that such coupled Eulerian-Lagrangian approach does in fact cover representation 
    of fractional cloudiness within a grid cell.
 
  \paragraph{DROPLET CONCENTRATION}

  The range of drop concentration values obtained in the simulations does roughly correspond to the
    values observed during RICO and presented in \citet[figure 1a]{Arabas_et_al_2009},
    \citet[figure~4]{Gerber_et_al_2008} and 
    \citet[figure~3]{Colon_Robles_et_al_2006}.
  The discrepancies in the median values and the spread of CDNC among different model runs show, however,
    that the prediction of drop concentration is sensitive to the super-droplet density (Fig.~\ref{fssp3}).
  Furthermore, the higher concentrations obtained with low super-droplet densities (i.e. low spectral resolution) 
    are closer to the observed values.
  The observed invariability of CDNC with height is robustly reproduced suggesting that the discrepancies
    are solely related to the treatment of CCN activation in the model.

  \paragraph{EFFECTIVE RADIUS}
  
  The effective radius profiles are generally robust to both grid and super-droplet density choices,
    and they do resemble the profiles observed with the Fast-FSSP instrument during RICO \citep[fig.~2]{Arabas_et_al_2009}.
  The profiles show a gradual increase in cloud droplet sizes from the cloud base up to the altitude of $700$--$800$ metres
    where the median values of $r_{e\!f\!f}$ reach $15$--$20$ micrometres.
  Above, the profiles differ more from one model run to another but still the flattening of the $r_{e\!f\!f}$ profile slope
    is a robust feature.
  The reduced slope of the median profiles of $r_{e\!f\!f}$ reflects (i) the chosen drop radius range 
    ($1$--$24$~$\mu$m -- emulating the Fast-FSSP instrument range), (ii) the decreased efficiency, 
    in terms of radius change, of the condensational growth for larger droplets, and  (iii) the increased 
    efficiency of drop collisions reached after the initial condensational growth stage \citep[see e.g. section
    3.1 in][for discussion of influence of precipitation on effective radius profiles]{Zhang_et_al_2011}.
  The spread of $r_{e\!f\!f}$ values high above cloud base indicates presence of smaller droplets even near cloud top.
  The present analysis (and the corresponding choice of data output rate during simulations) does not provide an answer to the
    question of the origin of these small particles -- e.g. if they were activated at cloud edges above cloud base
    or not \citep[see e.g.][]{Slawinska_et_al_2012}.

  \paragraph{LIQUID WATER CONTENT}

  LWC profiles depicted in Figures~\ref{fssp1}-\ref{fssp3} show significant spread of values
    at a given level.
  This is in accord with RICO observations, and suggests mixing-induced dilution of cloud and the resultant 
    decrease of water content \citep[cf. the discussion of the adiabatic fraction profiles in Fig.~2 in][]{Arabas_et_al_2009}.
  In SDM the condensational growth and evaporation is computed using 
    values of supersaturation interpolated to super-droplet positions, hence the mixing scenario is not homogeneous
    as different droplets are exposed to different supersaturations within a single grid cell.

  The decrease in LWC in the upper part of the cloud is correlated with an increase of rain water mixing
    ratio (not shown).
  In the lower part of the cloud, the median profiles of LWC obtained in the simulation with middle and high grid resolution 
    show arguably reasonable agreement with LWC profiles derived from RICO observations depicted in  
    \citet[][Fig.~8]{vanZanten_et_al_2011},
    \citet[][Fig.~1]{Gerber_et_al_2008} and 
    \citet[][Fig.~7]{Abel_et_al_2007}
    taking into account that the referenced observations were taken with 
    different instruments having different sampling rates and spectral ranges.

  \paragraph{PARAMETER K}

  Most GCMs employ a parameterisation of the effective radius when diagnosing the cloud optical depth 
    of clouds from predicted liquid water content and drop number concentration 
    \citep[see e.g. section 2 in][and references therein]{Brenguier_et_al_2000}.
  Typically, the cubed ratio of the mean volume radius to the effective radius $k\!\!=<\!\!r^3\!\!>/r_{e\!f\!f}^3$ 
    (or a similarly defined scaling coefficient) is assumed to be constant
    or to depend solely on drop concentration \citep[e.g.][]{Peng_et_al_2003}.
  Value of $k\!=\!0.8\pm0.07$ based on aircraft observations in maritime clouds using an 
    FSSP-type instrument reported in \citet{Martin_et_al_1994} is often used in climate models.

  As shown in Figure~\ref{fssp1}, all simulations with the mean value of 8 super-droplet per 
    LES grid cell predict values of $k$ hardly different from unity.
  This means that the effective radius is in most cases equal to the mean volume radius, as it 
    should be, taking into account that the 8 super-droplets (which may be thought of as size bins)
    are used to represent a wide spectral range of particles: aerosol, cloud, drizzle and rain; 
    while the statistics presented in the plot cover a narrow spectral range.
  Consequently, cloud water is likely represented by a single super-droplet (bin) in each grid cell, 
    a situation in which all of the cloud droplets would have the same size and $k=1$ by definition.
  The values of the $k$ parameter calculated in the model runs with mean super-droplet density of 128
    range from approximately 0.6 to 0.9.
  There is no significant difference when the super-droplet density is increased to 512 suggesting that 
    the density of 128 per cell is enough to resolve the relevant cloud droplet size spectrum features.

  \paragraph{DROP RADIUS STANDARD DEVIATION}

  In consistency with the aforementioned behaviour of $k$ for low super-droplet density,
    the drop radius standard deviation $\sigma_r$ hardly differs from zero when a mean of 
    8~super-droplets per cell is used, and it still goes down to zero at some levels 
    with a mean of 32 super-droplets per grid cell.
  For all higher spectral resolution simulations the 5\textsuperscript{th} percentile
    of $\sigma_r$ is greater than zero at all levels, and the increase of mean super-droplet
    density from 128 to 512 does not influence the profile shape.
  In general agreement with RICO observations, the standard deviation ranges from $1$ to $6$ micrometres.
  The profiles of $\sigma_r$ show a slight increase with height (best captured in the sdm-middle-128 run);
    however, the observed inclination of the $\sigma_r$ profiles resembles more the 95\textsuperscript{th}
    percentile profile derived from SDM simulations.
  The fact that the values of $\sigma_r$ are larger than those obtainable in adiabatic drop growth process 
    \citep[see e.g.][and references therein]{Yum_et_al_2005} suggests that
    the model does capture to some extent the mixing-induced broadening of the cloud droplet spectrum. 
  However, since the mixing in SDM is limited to LES-resolved motions with characteristic length scales constrained 
    by grid cell dimensions, the level of agreement with observations is, as expected, limited.

  \subsubsection{\!\!PRECIPITATION MICROSTRUCTURE}\label{sec:oap}
 
  The analysis depicted in Figures~\ref{fssp1}, \ref{fssp2} and \ref{fssp3} concerned the cloud-droplet
    region of the particle size spectrum.
  A comparison of model-predicted spectra for larger particles is depicted in Figure \ref{oap}.
  The figure is intended for comparison with Fig.~4 in \citet{Baker_et_al_2009} based on 
    measurement data obtained with the OAP-2DS instrument \citep{Lawson_et_al_2006} during RICO research flights.
  During RICO the OAP-2DS instrument was set to classify particles into 61 size bins spanning the 
    $2.5$~$\mu$m -- $1.5$~mm size range in radius.
  In the analysis of \citet{Baker_et_al_2009} a mean size spectrum was derived from 237 spectra measured 
    within rain-shafts below the cloud base at the altitude of about 183 metres (600 ft).
  In order to derive comparable quantities from the SDM simulation results, the super-droplets
    in each grid cell were classified into size bins of the same layout as used by the OAP-2DS instrument,
    an altitude range of $183\pm100$~m was chosen, and only grid cells with rain water mixing ratio
    $q_r>0.001$~g/kg were taken into account ($q_r$ being derived from summation over super-droplets
    representing particles with radii greater than $40$~$\mu$m).
  A comparison of Figure~\ref{oap} with Fig.~4 from \citet{Baker_et_al_2009} (both plots share
    the same axis ranges) reveal that the model results, regardless of the grid-resolution or super-droplet density choice, 
    show fair agreement with the measurement results for drop diameters greater than $0.1$~mm.
  The spectrum from the sdm-coarse-512 run (with the highest spectral resolution and the best Monte-Carlo sampling density) 
    resembles most closely the measurements having the lowest concentrations in the $0.1$--$0.4$~mm diameter range.
  All simulations disagree markedly with the measurements within the $10$--$20$~$\mu$m diameter range.
  These measured particles within this range were ''most likely deliquesced aerosols'' \citep{Baker_et_al_2009} and since
    there are no aerosol sources in the model, this disagreement is still a plausible result.
  The drop breakup process was identified as another possible source of droplets smaller than $100$~$\mu$m,
    and this process is also not included in the present version of the model.
  Moreover, it was the lack of hydrometeors smaller than $100$~$\mu$ in diameter that was considered
    as the primary highlight of the observations reported in \citet{Baker_et_al_2009}, and this feature 
    of the spectrum is in fact hinted in the simulation results.

  \subsection{SUMMARY}

  The salient features that distinguish SDM from bulk and bin warm-rain microphysics models are:\\ 
    (i) diffusive error-free computational scheme for both condensational and collisional growth;\\
    (ii) particle spectrum representation allowing straightforward comparison with experimental data obtained with particle-counting instruments;\\
    (iii) persistence of arbitrary number of scalar quantities assigned to a super-droplet (e.g. chemical properties);\\
    (iv) scalability in terms of sampling error (i.e.~super-droplet density);\\
    (v) parameterisation-free formulation of the key processes involved in cloud-aerosol interactions.

  The last feature, in particular, does not come without its challenges.
  Explicit treatment of aerosol microphysics implies taking care of their budget
    within the simulation domain, including modelling their sinks (wet and dry deposition) 
    and sources, the latter not being represented in the present study.

  The other two processes not represented in the discussed simulations are: (i) the 
    impact of turbulence on drop collisions \citep[see e.g.][]{Devenish_et_al_2011}, 
    and (ii) the influence of drop breakup on the size spectrum of precipitation particles
    \citep[see e.g.][]{Villermaux_et_al_2009}.
  Yet, the arguably fair level of agreement of the simulation results reported herein and the
    previously published in-situ measurement results from RICO, suggest that the set of
    processes represented in the present set-up of SDM includes at least the key players 
    involved in determining the features of the size spectra of cloud and precipitation particles.
  One has to bare in mind that a direct comparison of RICO measurement data with LES results 
    of the type presented herein
    is not possible due to different time- and space- scales associated with the model variables
    and the measurements, as well as due to the nature of the "composite" modelling set-up.
    (typical atmospheric conditions, with the diurnal cycle neglected in particular).

  The SDM modelling approach offers the unique possibility to mimic in the analysis 
    the particle-counting process inherent in the principle of operation of most airborne aerosol, 
    cloud and precipitation probes.
  As a result, it becomes meaningful to analyse such model-predicted cloud droplet size spectrum
    parameters as e.g. $k$ or $\sigma_r$, not taken into account in previous comparisons
    of RICO LES results with aircraft observations \citep{Abel_et_al_2007,vanZanten_et_al_2011}.
  What militates in favour of pertinence of the obtained results is that the present model 
    employs fewer parameterisation (in comparison with bulk or bin models) and more basic 
    principles to describe processes occurring at the microscale
    (e.g. description of the K\"ohler curve shape as opposed to employment of such parameters
       as the exponent in Twomey's formula for CCN activation parameterisation in bin models, 
       or the autoconversion threshold in Kessler-type bulk models).

  The key conclusion from the presented analysis is that while the convergence of the macroscopic
    cloud parameters do not seem to get any better than in the other LES simulations using the RICO
    set-up, the SDM is able to provide more detailed insight into cloud microstructure, and
    thus indirectly into its optical properties.

  \linespread{.85}
  \scriptsize

  \vspace{.3cm}
  {\bf\noindent\normalsize Acknowledgements}\\
  ~\\
  All simulations were carried out on ''The Earth Simulator~2'' operated by the 
    Japanese Agency for Marine-Earth Science and Technology (JAMSTEC) in Kanagawa, Japan.
  JAMSTEC supported a month-long research visit of SA to Japan, and provided computer time on the ES2.
  Thanks are due Hanna Pawlowska (University of Warsaw) and Kanya Kusano (JAMSTEC, Nagoya University) 
    for their support throughout the project; and Kozo Nakamura (JAMSTEC) for his help with implementing 
    the RICO set-up in CReSS.

  \vspace{.3cm}
  {\bf\noindent\normalsize References}\\
  ~\\
  \bibliographystyle{copernicus}
  \vspace{-1.5cm}
  \bibliography{ricosdm}

\end{document}